\documentclass[11pt,a4paper]{article} 

\usepackage{jcappub}
\usepackage{bm}

\def\a{\alpha}
\def\b{\beta}

\def\e{\epsilon}

\def\L{\Lambda}

\def\O{\Omega}
\def\th{\theta}

\def\s{\sigma}
\def\ra{\rightarrow}

\title{Smooth halos in the cosmic web}

\author{Jos\'e Gaite} \affiliation{Physics Dept., ETSIAE, IDR, Universidad
  Polit\'ecnica de Madrid, E-28040 Madrid, Spain}

\emailAdd{jose.gaite@upm.es}

\abstract{ Dark matter halos can be defined as smooth distributions of dark
  matter placed in a non-smooth cosmic web structure. This definition of halos
  demands a precise definition of smoothness and a characterization of the
  manner in which the transition from smooth halos to the cosmic web takes
  place. We introduce entropic measures of smoothness, related to measures of
  inequality previously used in economy and with the advantage of being
  connected with standard methods of multifractal analysis already used for
  characterizing the cosmic web structure in cold dark matter $N$-body
  simulations. These entropic measures provide us with a quantitative
  description of the transition from the small scales portrayed as a
  distribution of halos to the larger scales portrayed as a cosmic web and,
  therefore, allow us to assign definite sizes to halos. However, these
  ``smoothness sizes'' have no direct relation to the virial radii. Finally,
  we discuss the influence of $N$-body discreteness parameters on smoothness.}


\toccontinuoustrue

\begin{document}

\maketitle

\section{Introduction}
\label{intro}

The large scale structure of the Universe can be described as a ``cosmic web''
formed by matter sheets, filaments, and nodes, plus the cosmic voids that
these objects leave in between.  This type of structure was initially proposed
in connection with simplified but insightful models of the cosmic dynamics,
namely, the Zeldovich approximation and the adhesion model
\cite{Zel,Gurb-Sai,Shan-Zel,GSS}. It has been since confirmed by galaxy
surveys \cite{EJS,ZES,Ge-Hu} and cosmological $N$-body simulations
\cite{Kof-Pog-Sh,WG,Kof-Pog-Sh-M}.  $N$-body simulations have especially
contributed to the understanding of structure formation.  In particular, the
analysis of cold dark matter (CDM) $N$-body simulations has consistently shown
that, on smaller scales, the cosmic-web structure transforms into a
distribution of relatively smooth dark matter clusters or {\em halos} that
have a limited range of sizes.  Halo models of the large scale structure of
matter \cite{CooSh} are now very popular indeed.  Dark matter halos were
initially introduced to model the invisible matter surrounding galaxies, but
present halo models are concerned with the large scale distribution of halos
in space \cite{MoW,CMP,Sheth-Tor} as well as with the distribution of matter
within individual halos \cite{Power,Hayashi,Navarro}.  Actually, the essence
of halo models is to separate the full dark matter distribution into one part
corresponding to the distribution of dark matter inside halos and another
corresponding to the distribution of halos centers in space \cite{CooSh}.  The
distribution of dark matter inside a halo is smooth, save for the density
singularity at its center and the possible presence of other halos (subhalos)
close to it.  The distribution of halo centers in space must follow the
cosmic-web structure.

The geometry of the cosmic-web structure in the adhesion model belongs to the
geometric type of mass distributions that have noticeable geometric features
on ever decreasing scales.  This type of geometry is generally referred to as
fractal geometry \cite{Mandel}.  Fractal geometry typically appears in
nonlinear dynamical systems in which the dynamics is characterized by the
absence of reference scales and is driven to an attractor, independent of the
initial conditions.  The dynamics of collision-less CDM, only ruled by the
gravitational interaction, is scale invariant, although the initial conditions
are not and it is usually assumed that they determine the geometry in the
nonlinear regime.  The adhesion model \cite{Shan-Zel,GSS} gives rise to a
self-similar cosmic web that indeed depends on an initial \emph{power-law}
spectrum of fluctuations \cite{V-Frisch}; but in the \emph{stochastic}
adhesion model \cite{Kturb,Rigo}, equivalent to the Kardar-Parisi-Zhang
equation of interface growth \cite{Bou-M-Parisi}, the cosmic-web is actually
an attractor independent of the initial conditions.  At any rate, the fractal
analysis and, more specifically, the multifractal analysis of the large-scale
structure have a long history, including analyses of the distribution of
galaxies \cite{Pietronero,Jones,Bal-Schaf,Jones-RMP} and of CDM $N$-body
simulations \cite{Valda,Colom,Yepes,fhalos,I4,voids,MN}.  Furthermore, the CDM
structure produced by $N$-body simulations can be described as a distribution
of halos in a multifractal cosmic structure \cite{fhalos,I4,ChaCa}.  Whether
or not the cosmic web is self-similar, its multifractal spectrum and,
specifically, its R\'enyi dimensions can be reliably computed in CDM $N$-body
simulations (multifractality does not imply self-similarity) \cite{I4,MN}.
That is what we need in the present work.

The adhesion model is soluble and produces a distribution of singular sheets,
filaments and nodes of vanishing size in the limit of vanishing viscosity
\cite{Shan-Zel,GSS}. The regularizing effect of a finite viscosity smoothens
these structures and gives them a size proportional to it.  This suggests that
the smoothness of halos in CDM $N$-body simulations may be influenced by the
regularizing effects, on small scales, of $N$-body discreteness and the
associated gravity softening \cite{I6}. In other words, the range of halo
sizes may depend on the discreteness scales, namely, the discretization length
$N^{-1/3}$ (length of the cube with one particle on average), and the
gravity-softening length. $N$-body discreteness primarily affects underdense
regions: the structure of cosmic voids is lost on scales smaller than the
discretization length \cite{I4,voids}.  Therefore, the web structure must
undergo a morphological transition on scales of the order of the
discretization length, and smaller scales can only provide, at best, a
distorted portrait of the cosmic web.

At any rate, halos, as high density regions, are generally well sampled below
the discretization length, unlike voids.  In fact, the part of the cosmic-web
multifractal spectrum that corresponds to high density regions and,
consequently, to halos can be obtained accurately on scales considerably
smaller than $N^{-1/3}$ \cite{I4,MN}.  Therefore, one may wonder why the
smooth aspect of halos is so different from a cosmic-web structure, that is to
say, why such a drastic transformation takes place on scales close to the
discretization length.  A thorough analysis of the influence of discretization
on the sizes and the smoothness of halos would require us to compare various
$N$-body simulations and, therefore, it would demand a considerable use of
computing resources. Before undertaking this job, it is necessary to have a
better understanding of the factors determining the size and smoothness of
halos and the transition to the cosmic web structure in $N$-body
simulations. This can be achieved with the analysis of one $N$-body simulation
with good resolution. We analyze the Bolshoi simulation \cite{Bolshoi,MD}.

Dark matter halos were initially conceived as dark matter concentrations that
are approximately spherical and centered on peaks of the density field,
although now it is understood that halos are more or less ellipsoidal
\cite{CooSh}.  Halos are usually bounded at their virial radii, but there is
no natural halo boundary and there are various definitions of it
\cite{halo-find}.  The definition that places the halo boundary at the virial
radius can be criticized on various grounds and, especially, concerning the
suitability of the spherical collapse model \cite{I6}.  An alternative is
precisely to use smoothness as the property of the dark matter distribution in
a halo that defines its boundary.  One of the questions we intend to answer is
whether the ``smoothness'' radius is related to the virial radius.

Although smoothness is easily perceived by the human eye, a precise
(mathematical) definition of it is not obvious and may depend somewhat on the
application. Therefore, our first concern must be to provide a
characterization of smoothness that is suitable for $N$-body simulations.  Of
course, the smoothness of dark matter halos, or, rather, their non-smoothness,
has already been studied in the literature. In fact, one of the central
problems of the CDM model, namely, the ``missing satellites problem'', is
directly related to the graininess of dark matter halos \cite{CDM_PNAS}.  To
quantify the graininess of dark matter halos, Zemp et al \cite{Zemp} employ
statistical measures and apply them to a Milky Way-mass dark matter halo in an
$N$-body simulation, namely, the Via Lactea II (VL2) simulation \cite{VL2}.
Our work is related to Zemp et al's \cite{Zemp}, but we consider the question
of inner halo structure in relation to the larger scale structure, that is to
say, our concern is the transition from a smooth distribution on halo scales
to a non-smooth and strongly anisotropic cosmic-web structure on larger scales
(or vice versa).  Therefore, our statistical methods are essentially different
from theirs and actually are an adaptation of multifractal methods to the
analysis of individual halos.  After developing this method, we can analyze
the smoothness of halos in $N$-body simulations to determine the variation of
smoothness with growing halo radius and determine how smoothness disappears
and gives way to cosmic-web non-smoothness.

As mentioned above, $N$-body discreteness effects must play a role in the
transition from smooth halos to the cosmic web in CDM simulations.
Discreteness effects are due to having $N$ bodies and the related gravity
softening.  A softening length is needed in every method of gravity softening
and is commonly chosen to be much smaller than the discretization length
$N^{-1/3}$ \cite{DR}.  Whether this is correct or not is a controversial issue
\cite{KMS,SMSS,Romeo,JMB}, but it is commonly accepted that it is.  We briefly
study the influence of the two scales on the size of halos, which is an
important but still moot question.

Last, let us mention, as a matter of interest, that some new studies of cosmic
structure consider the description of structure in the six-dimensional phase
space.  Zemp et al \cite{Zemp} already relate graininess of the spatial
distribution to features of the velocity field that can be interpreted as the
presence of streams of matter. The multi-stream nature of phase space is
further studied by Shandarin \cite{Shanda-JCAP}, Abel et al \cite{Abel} and
Neyrinck \cite{Ney}. Our method for the analysis of smoothness of the
distribution in real space can be extended to phase space, but this extension
is beyond the scope of the present work.

To summarize, our plan is the following.  The problem of halo smoothness is
presented in Sect.~\ref{smooth}. We introduce an entropic measure of
(non)smoothness that is suitable for $N$-body simulations and constitutes a
new method of multifractal analysis of halos.  Then, we compare our entropic
measure to the measures of Zemp et al (Sects.~\ref{square} and
\ref{ent-mea}). In Sect.~\ref{Nbody}, we apply our measure to a number of
halos from the Bolshoi simulation and to the Milky-Way VL2 halo.  The latter
is useful for a quantitative comparison with the results of Zemp et al, but
the Bolshoi simulation contains a large number of halos which allow us to
compare different halos in the same simulation.  In addition, the Bolshoi
simulation is suitable for a multifractal analysis of the cosmic-web structure
(Sect.~\ref{MF-Bolshoi}) that is useful to characterize the transition from
halos to the cosmic web. In Sect.~\ref{discret}, we consider the relation
between the smoothness of halos and the discreteness parameters of $N$-body
simulations.  We present our conclusions in Sect.~\ref{discuss}.  Finally, we
include appendix \ref{append}, with basic techniques of multifractal analysis,
as applied to $N$-body simulations.

\section{Smoothness and isotropy of halos}
\label{smooth}

A dark matter halo consists of a distribution of dark matter particles with a
radial density profile that is singular at the center \cite{CooSh}.  In
general, this singularity is found to be of power-law type, although its exact
form could be slightly more complicated \cite{Navarro}.  For $r>0$, the
density is finite and a smooth function of the coordinates.  The question
addressed in this paper is the extent of the smooth distribution of matter
that can be associated with halos rather than the precise properties of the
halo radial density profile.  The question of smoothness is essentially the
same question studied by Zemp et al \cite{Zemp}, because graininess is
opposite to smoothness, so smoothness ends where graininess begins. Although
it may be taken for granted that the distribution of dark matter is smooth on
sufficiently small scales, this is not a logical necessity, and in a fully
multifractal cosmic web structure the singularities appear everywhere, not
only in isolated halo centers \cite{fhalos,I4}.

\begin{figure}
\centering{\includegraphics[width=7.4cm]{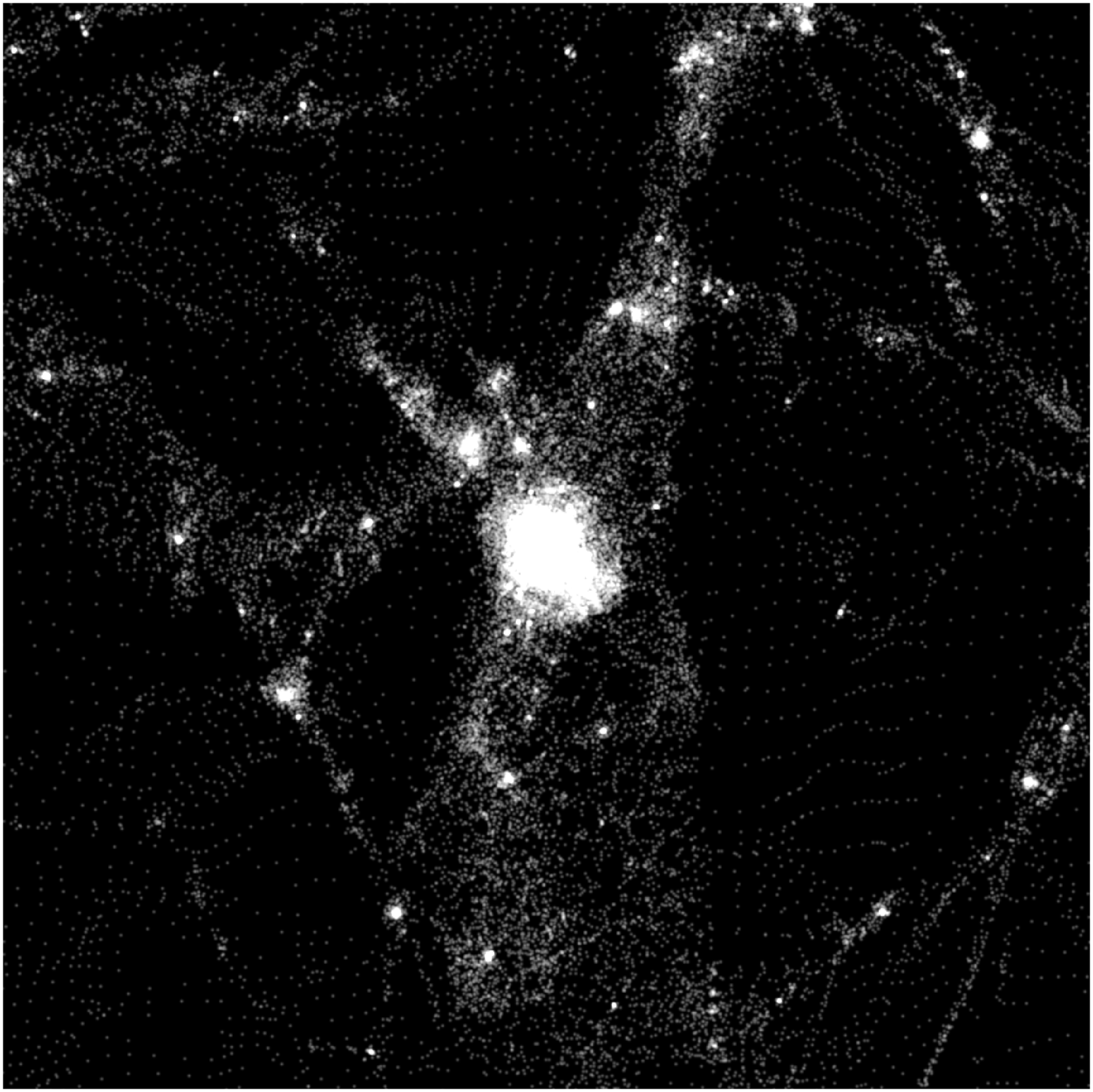}
\hspace{3mm}
\includegraphics[width=7.4cm]{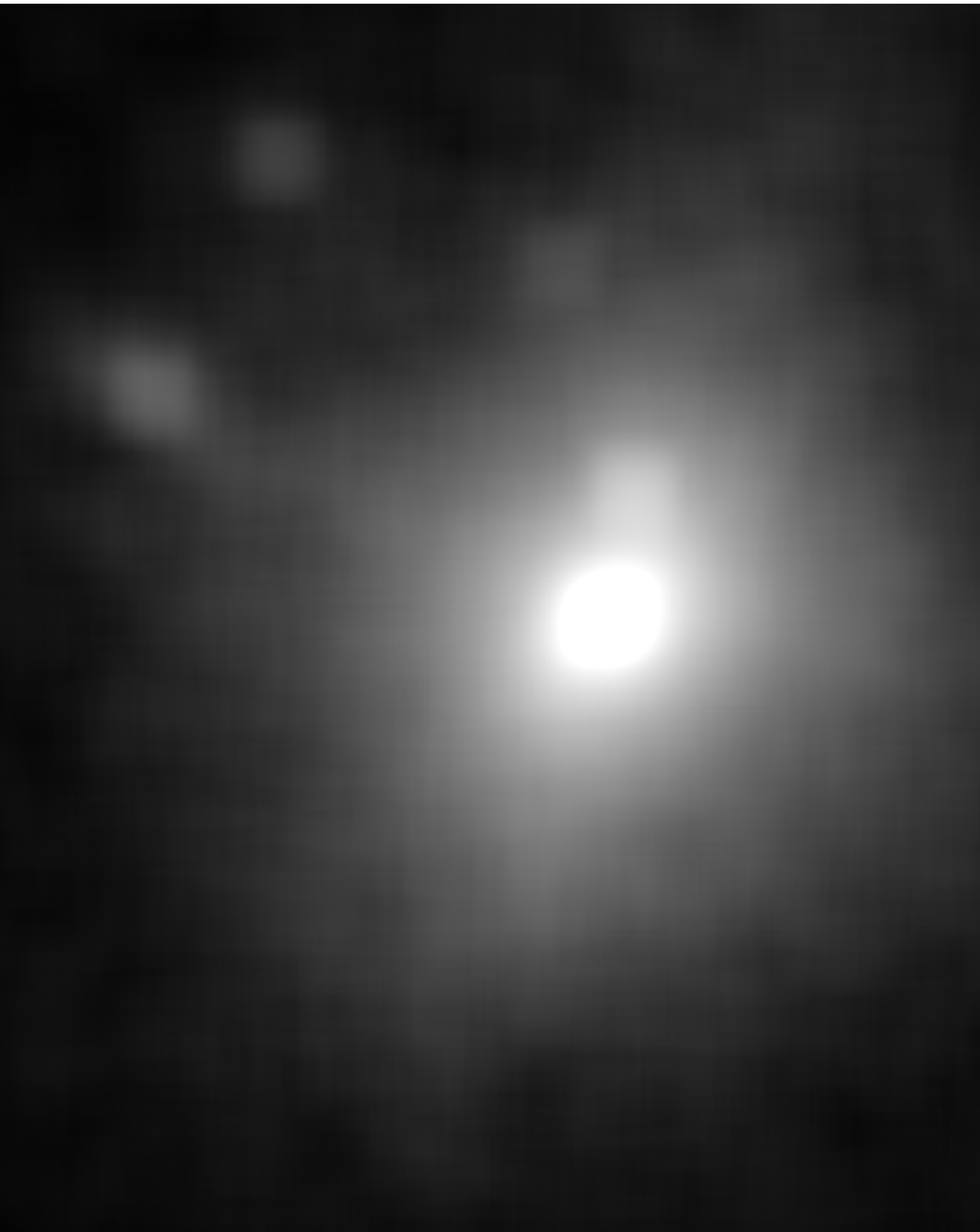}
}
\caption{(Left) Cosmic web around a Bolshoi simulation massive halo,
  at the scale of 15.6 Mpc$/h$, showing a grainy and filamentary structure.
(Right) Close-up on the same halo, spanning 458 kpc$/h$, showing a smooth and 
quasi-spherical mass concentration.}
\label{halo-zoom}
\end{figure}

The first problem to characterize the smoothness of halos is that there is no
general agreement on how to define individual halos in $N$-body simulations
(for a recent and comprehensive reference about halo finding, see
\cite{halo-find}). Nevertheless, halos certainly are mass concentrations, and
most halo finders begin by locating peaks in a suitably defined coarse-grained
density field, which are the potential halo centers \cite{halo-find}. Of
course, there is no unique definition of this coarse-grained density field, so
the locations of halo centers may be slightly inaccurate, but this is not
important. Once chosen one halo center, it is necessary to determine the
extent of the halo. According to our hypothesis, the halo ends where the
smooth distribution of particles transforms into a grainy distribution
identifiable with the expected large-scale cosmic-web structure.  This
transformation is obvious as we zoom in or out on any halo, as can be seen in
Fig.\ \ref{halo-zoom} (where the density field has been obtained by Gaussian
filtering with $\s=15$ kpc$/h$).%
\footnote{Cosmic web features are clearly visible in the density field, but
  sharper renderings of them are provided by Abel et al's visualization method
  \cite{Abel}, which takes advantage of the full phase space structure.}  
If we imagine a spherical surface with origin on a halo center and with
increasing radius, at some radius, that surface must intersect a distribution
that is essentially indistinguishable from the cosmic web.

Let us consider the cosmic web produced by the adhesion model, which is
calculable and reaches infinitesimally small scales \cite{V-Frisch}.  This
cosmic-web structure displays strong an\-isotropy and a rapid variation of the
(coarse-grained) density between neighboring points, unlike a smooth matter
distribution.  The anisotropy and the rapid variation of density are also
perceived in images of $N$-body simulations, e.g., Fig.\ \ref{halo-zoom},
left. But we need precise mathematical definitions that allow us to measure
them.  Mathematically, a function is smooth if it is
differentiable. Therefore, the natural procedure to determine the smoothness
of a point distribution that results from an $N$-body simulation should be to
compute derivatives of the coarse-grained density. Thus, the first problem
would be to define a coarse-grained density and its derivatives. However,
there arises a serious problem with this method: one cannot expect smoothness
of the small scale dark matter distribution in the mathematical sense, because
halo radial density profiles are singular at $r=0$. At a singular point, the
density and its derivatives diverge.  Nevertheless, we expect that isolated
singularities preserve some degree of smoothness, unlike the singularities in
typical multifractals, which make them totally non-smooth.  However, we have
to bear in mind that a collection of isolated power-law singularities can
approach an ordinary multifractal as the density of singular points increases
\cite{fhalos}, so the difference between a multifractal cosmic web and a
suitable distribution of smooth halos with power-law profiles is quantitative
rather than qualitative.  We can measure the degree of smoothness by comparing
global measures of the magnitudes of the derivatives of the coarse-grained
density as functions of the coarse-graining length.  We have tried this method
but it is rather cumbersome.

At any rate, a more useful feature distinguishing between the dark matter
distribution inside halos and the cosmic web is the change from mild to strong
anisotropy.  A deep study of the cosmic-web anisotropy requires sophisticated
mathematical concepts. Indeed, the cosmic-web anisotropy is due to its
morphological structure, namely, to its filaments and sheets, which are
described through sophisticated topological constructions \cite{S3,S4,vW-Sch}.
Whatever the method employed to measure the cosmic-web morphology, we know
that the anisotropy, close to a halo center, can be just reduced to an
ellipsoidal profile, whereas, on larger scales, the anisotropy patterns are
much more complex (Fig.\ \ref{halo-zoom}). If we consider a spherical shell
centered on a halo, the density in it must be fairly smooth for small radius
but become increasingly non-smooth as its radius grows. In other words,
non-smoothness and anisotropy are closely related. To be precise, the type of
anisotropy that is relevant for delimiting a halo is characterized by the
degree of non-smoothness inside spherical shells.  Therefore, smoothness can
be measured through anisotropy and vice versa.  Note that the singularity at
the halo center is avoided by measuring smoothness only in spherical shells.

Now, we need to devise a measure of non-smoothness of the distribution in a
spherical shell.  We expect to find a smooth source of anisotropy at all
radii, namely, triaxiality, but also sharp local deviations from isotropy due
to the presence of subhalos, which must appear at some distance from the
center. Farther from the center, there appear more complex patterns, in
particular, underdense regions and holes \cite{Zemp}, as the shell intersects
the voids of the cosmic web structure. In summary, we must expect that the
smooth distribution in a small-radius shell transforms, as the radius gets
large, into a very inhomogeneous distribution that displays features
corresponding to the cosmic web.

\subsection{Counts in cells for spherical shells}
\label{square}

To measure the non-uniformity of the distribution of particles in a spherical
shell, we can compare the angular spherical coordinates of the particles in
it, say $\{\phi_i,\th_i\}_{i=1}^k$, to the ones corresponding to a uniform
distribution. To do this, we must realize that, for a uniform distribution in
spherical coordinates, the azimuth angle $\phi$ must have a uniform
distribution in $[0,2\pi]$, but this does not apply to the polar angle $\th$:
actually, it is $\cos\th$ the quantity that must be uniformly distributed in
its range, $[-1,1]$. For simplicity, we use $\phi/(2\pi)$ and $(1+\cos\th)/2$,
which must have a uniform density in the unit square $[0,1] \times [0,1]$.
These coordinates, of course, belong to the set of coordinate systems defined
in geography for cylindrical equal-area projections \cite{mathworld} (but note
that the square projection is uncommon). To check if the density of the $k$
values of these redefined angular coordinates conforms to uniformity, we can
use counts in cells: in a uniform distribution, the fluctuations of counts in
cells conform to the binomial distribution or, for sufficiently large samples,
to the Poisson distribution.

While it is easy to test for the Poisson distribution, for example, we must
take into account that there are two expected sources of non-uniformity,
namely, the triaxiality of halos and the presence of subhalos in a given halo.
This is pointed out by Zemp et al \cite{Zemp}, who propose to evaluate
underdense regions in the VL2 halo as a more suitable measure of its
graininess.  In this regard, one could apply to the study of voids in the
distribution of particles in a spherical shell statistical methods similar to
the ones applied to voids in the full three-dimensional particle distribution
(such as the methods in \cite{voids}, for example).  Zemp et al \cite{Zemp}
actually use one elementary statistic: the void probability function, namely,
the probability that a {\em given} region be empty (the region that they take
is a small ball with center in the shell). This statistic is sufficient to
rule out (for large enough shell radius) a uniform distribution or even a
smooth triaxial distribution with subhalos, as they do.

The void probability function can also be estimated from counts-in-cells in a
spherical shell.  However, it is useful to consider quantities that are not
only concerned with almost empty cells, belonging to voids.  In fact, the most
useful quantities must take all the cells into account and provide a measure
of the inequality or statistical dispersion of counts-in-cells.  Such
quantities are commonly employed in economy, for example, to measure
inequality of income, where the units of income are individuals, cities, etc.\
\cite{Ul}.  Zemp et al \cite{Zemp} employ the Gini coefficient, one inequality
measure that has become very popular. Another inequality measure very popular
in economy is Theil's entropic index \cite{Ul}, which is inspired by
information theory. Indeed, the problem of income distribution is just one
instance of the general problem of the partition of some measurable quantity
(mass, money, etc.). When this quantity is discrete, the partition problem is
equivalent to the combinatorial problem of the distribution of a set of
particles in a number of cells. The standard measure of uncertainty in the
choice of one particle is the Boltzmann-Gibbs-Shannon (BGS) entropy. A useful
generalization of it is the R\'enyi entropy \cite{Renyi}, which constitutes a
suitable measure of the statistical dispersion of a partition in cells and, as
such, can be used in the analysis of cosmological $N$-body simulations
\cite{MN}. Here, this dispersion measure is applied to spherical shells of a
halo.  Let us notice that generalized entropy indices are also employed in
economy \cite{Ul}, but they are based on a type of entropy that is different
from R\'enyi entropy and does not have all its desirable properties; in
particular, it is not additive.

\subsection{Entropic measures}
\label{ent-mea}

The R\'enyi entropies
\begin{equation}
S_q(\{p_i\}) = \frac{\log_2 (\sum_{i=1}^M {p_i}^q)}{1-q}, \quad 
q \neq 1\,,
\label{Sq}
\end{equation}
measure the statistical dispersion of counts-in-cells $\{n_i\}_{i=1}^M\,,$
corresponding to the partition of $N=\sum_{i=1}^M n_i\,$ particles in $M$
cells, in terms of ``probabilities'' $\{p_i = n_i/N\}_{i=1}^M$.  The limit of
$S_q$ as $q \to 1$ just yields the standard BGS entropy.  The R\'enyi
entropies with $q \geq 0$ are bound, namely, $0 \leq S_q(\{p_i\}) \leq \log_2
M$.  Hence, it is convenient to divide them by $\log_2 M$, so that they become
numbers between 0 and 1, the former corresponding to maximum order or
inequality, and the latter to minimum order (uniformity). Then, these bounds
are the same ones as the bounds of the Gini coefficient, although their
meaning is reversed. The entropic coefficients defined in that way, namely,
$S_q(\{p_i\})/\log_2 M$, are related to R\'enyi \emph{dimensions} (see
appendix \ref{append}). Indeed, the entropic coefficients are just coarse
R\'enyi dimensions \cite{Falcon} divided by the dimension of the ambient
space, which is, in the present case, a two-dimensional spherical surface,
instead of the ordinary three-dimensional Euclidean space.  Therefore, we can
consider the entropic coefficients alternately as constituting a particular
type of inequality measures or as a sort of coarse R\'enyi dimensions
(independent of the dimension of the ambient space).

The connection of entropic coefficients with R\'enyi dimensions solves the
general problem of the dependence of inequality measures on the chosen unit,
that is to say, on the size of the cell, in our case.  Unlike in economy,
where the division of income into individuals or other units is natural, our
cell size is arbitrary. However, this arbitrariness is immaterial provided
that the coarse R\'enyi dimensions converge to their values $D_q$ in the
continuum limit, namely, in the limit in which the number of particles and the
number of cells tend to infinity.  This convergence takes place in
multifractals and guarantees certain independence of cell size.  Indeed, a
multifractal distribution is precisely defined by the existence of the moment
exponents $\tau(q)=(q-1)D_q\,.$ Notice that self-similarity is a sufficient
but \emph{not necessary} condition for it.  The property of the R\'enyi
entropic coefficients of converging to definite values in the continuum limit
is not shared by the Gini coefficient or other inequality measures.  The
coarse R\'enyi dimensions of halo shells, in addition to being only mildly
dependent on the cell size, are certainly useful to relate individual halos to
the full multifractal cosmic-web structure.

Therefore, we employ the entropic coefficients $S_q(\{p_i\})/\log_2 M$ of the
counts-in-cells in the unit square corresponding to equal-area angular
coordinates of particles in a spherical shell of a halo.  Next, we have to
determine the thicknesses of shells and the numbers of cells in them. These
are related issues: every shell must contain a sufficient number of particles
for meaningful counts, that is to say, for not having too small numbers of
cells and of particles per cell.  A too small number of cells may average out
large fluctuations that take place on small scales. Indeed, the entropic
coefficients $S_q/\log_2 M$ are certain to approach the R\'enyi dimensions
only for large $M$.  On the other hand, for a given number of particles in one
shell, a too large $M$ leaves most cells empty, and the occupied ones can only
have too small numbers of particles. Unfortunately, we cannot have a big
number of particles per shell, especially, for small radii, because it makes
the shell too thick; that is to say, a compromise is needed. We find it
suitable to have $1024$ particles per shell, for any radius, and $M=64$ cells
per shell, obtained with a $8 \times 8$ mesh in the unit square.  These 64
cells play the role of the $10^4$ spheres used by Zemp et al \cite{Zemp} for
the computation of Gini coefficients (and other quantities). Our number of
cells is much smaller but is sufficient: we have checked that it does not lead
to noticeable statistical errors (for reasonable values of $q$).

When we observe how the $q \geq 1$ entropic coefficients of a shell in a
given halo vary with the radius of the shell, we notice that they start at
values close to one and have a generally decreasing trend. This is in accord
with the expected transition from small-scale smoothness to large-scale
graininess. However, we can also notice that the regular decreasing trend is
punctuated by sudden dips, which naturally correspond to strong
inhomogeneities due to subhalos. Subhalo singularities strongly alter an
otherwise fairly smooth distribution. Of course, this expected source of
non-uniformity is best discarded. In a thin shell, there can only appear
one subhalo, or perhaps a few of them. To avoid taking them into
account in the computation of entropic coefficients, we can just remove a few
of the most populated cells of every shell. This hardly alters the overall
smoothness properties of the distribution in the shell but avoids subhalo
singularities. We choose to remove the four most populated cells of every
shell, reducing $M$ to 60. Therefore, the entropic coefficients are given by
$$\frac{S_q\left(\{n_i\}_{i=1}^{60}\right)}{\log_2 60} = \frac{S_q}{5.907}\,.$$
This simple operation reduces substantially the disturbing effects of
subhalos.

Our complete procedure consists of splitting a halo in successive shells with
$1024$ particles each, in an onion-like structure, and computing a number of
entropic coefficients for each shell, up to values of the radius such that the
coefficients stabilize (if it so happens). The $q \geq 1$ coefficients must
always decrease outwards. Assuming that the $q=0$ R\'enyi dimension of the
cosmic web is $D_0=3$, then the $q=0$ entropic coefficient must be always
close to one and the $q > 0$ coefficients must decrease outwards. Conversely,
the $q < 0$ coefficients should increase outwards. For sufficiently large
radii, all the coefficients must approach the ones that correspond to the
cosmic web.  The variation of entropic coefficients with radius must display a
gradual transition from smoothness to graininess (except for the local effects
of subhalos). This transition is similar to the one found by Zemp et al
\cite{Zemp}.

In the next section, we consider the specific contribution of triaxiality as a
smooth source of inequality of counts-in-cells.

\subsection{Graininess versus triaxiality}

The density in a spherical shell of a triaxial halo is not uniform, so the
entropic coefficients can deviate from one even in the absence of real
graininess. If the mass resolution of an $N$-body simulation halo were such
that we could have a very large number of particles per shell, we could easily
differentiate triaxiality from real graininess by substantially increasing the
number of cells, $M$, which would make irrelevant any smooth variation of
density, in particular, variations due to triaxiality.  Indeed, by increasing
$M$, we would be approaching the computation of R\'enyi dimensions, which are
unaffected by any smooth variation of density. However, with only $1024$
particles per shell and $M=64$, we need to estimate the effect of triaxiality.

To see how anisotropy due to triaxiality but not to graininess reflects on the
entropic coefficients computed with $1024$ particles in $64$ cells, we
calculate these coefficients for a smooth distribution with considerable
triaxiality, namely, a density with a deformed power-law radial profile: the
density with profile $r^{-2}$ subjected to an affine transformation to obtain
axis ratios 2/3 and 1/3. Our smoothness measuring procedure, applied to 1024
points in a 0.4\%-thick shell, yields coefficients 1, 0.965, 0.940, for
$q=0,1,2,$ respectively.  The last two coefficients differ significantly from
one, yet they are close to one.%
\footnote{The $q=0$ coefficient is smaller than 1 if one cell, at leat, is
  empty (this coefficient is related to the void probability function
  \cite{voids}). But, if there are few empty cells, the difference is
  negligible.}  However, there are many Bolshoi halos with ratios of minor to
major axis smaller than 1/3; and, in fact, there are even large halos with
ratios close to 1/10.  To prevent errors in the computation of entropic
coefficients due to the effect of triaxiality combined with insufficient mass
resolution, we may select quasi-spherical halos (with ratios of minor to major
axis close to one). At any rate, it is worthwhile to examine some strongly
triaxial halos for a comparison.

\section{Smoothness of halos in $N$-body simulations}
\label{Nbody}
 
Now we analyze the smoothness of halos in the Bolshoi and VL2 simulations,
with the procedure described above. We first summarize the characteristics of
these simulations. Furthermore, since the multifractal properties of the
cosmic web play a role in our arguments, we also provide the results of a
multifractal analysis of the Bolshoi simulation, carried out as explained in
appendix \ref{append} (which is based on the techniques employed in \cite{I4}
and, especially, in \cite{MN}).

\subsection{Bolshoi and Via Lactea II simulations}

The Bolshoi $\L$CDM simulation is described by Klypin et al
\cite{Bolshoi}. Here we quote its most relevant parameters. The simulation
assumes cosmological parameters $\O_\L=0.73$, $\O_{\rm M}=0.27$, $\O_{\rm
  bar}=0.0469$, Hubble parameter $h = 0.70$, and initial spectral index
$n=0.95$.  The edge length of the (comoving) simulation box is 250 $h^{-1}$
Mpc and the number of particles $N=2048^3$, which amounts to a mass resolution
of $1.35\cdot 10^8\:h^{-1}\, M_\odot$ per particle and a discretization length
of $0.122$ $h^{-1}$ Mpc.  The (Plummer) softening length is 1 $h^{-1}$ kpc
(physical, that is, not comoving).  Our statistical analysis only requires the
present time $z=0$ snapshot and the corresponding list of halos, both obtained
from the MultiDark database \cite{MD}. Naturally, we are interested in the
bound-density-maxima (BDM) halos rather than in the friends-of-friends (FOF)
halos (like in \cite{Bolshoi}, where there is information about the latter as
well).

The VL2 simulation \cite{VL2} focuses on the formation of a single, Milky-Way
size CDM halo, using the method of refinement. This simulation assumes
cosmological parameters $\O_\L=0.76$, $\O_{\rm M}=0.24$, Hubble parameter $h =
0.73$, and initial spectral index $n=0.95$. The edge length of
the (comoving) simulation box is 40 $h^{-1}$ Mpc. The halo is refined with
more than $10^9$ high-resolution particles, achieving a resolution of $4098\,
M_\odot$ per particle.  The softening length is 40 pc (physical after $z=9$).

\subsection{Multifractal analysis of the Bolshoi simulation}
\label{MF-Bolshoi}

The multifractal analysis of the Bolshoi simulation is carried out using the
counts-in-cells method described in detail in appendix \ref{append}.  In the
multifractal analysis of an $N$-body simulation by counts-in-cells, there are
two scales that play a fundamental role: the homogeneity scale and the
discretization length $N^{-1/3}$. The former is a physical scale, produced by
the evolution of gravitational clustering, whereas the latter is intrinsic and
indicates the scale at which the discretization effects dominate, on average.
The multifractal cosmic-web structure must appear between those two scales.
The homogeneity scale of the Bolshoi simulation, determined as explained in
appendix \ref{append}, is $l_0 = 15.6$ $h^{-1}$ Mpc, similar to the values
found before in the GIF2 and Mare-Nostrum simulations \cite{I4,MN}.%
\footnote{Remarkably, it is \emph{exactly} the same value as in the
  Mare-Nostrum simulation \cite{MN}. This coincidence is due to our using cell
  sizes that are powers-of-two fractions of the simulation box edge, and to
  the box edge of the Bolshoi simulation being precisely one half of the box
  edge of the Mare-Nostrum simulation.  } Actually, the transition to
homogeneity is not very sharp, beginning at a scale of about $8$ $h^{-1}$ Mpc
and ending at about $30$ $h^{-1}$ Mpc.  The discretization length,
$N^{-1/3}=2^{-11}$, is 0.12 $h^{-1}$ Mpc.

\begin{figure}
\centering{\includegraphics[width=7.4cm]{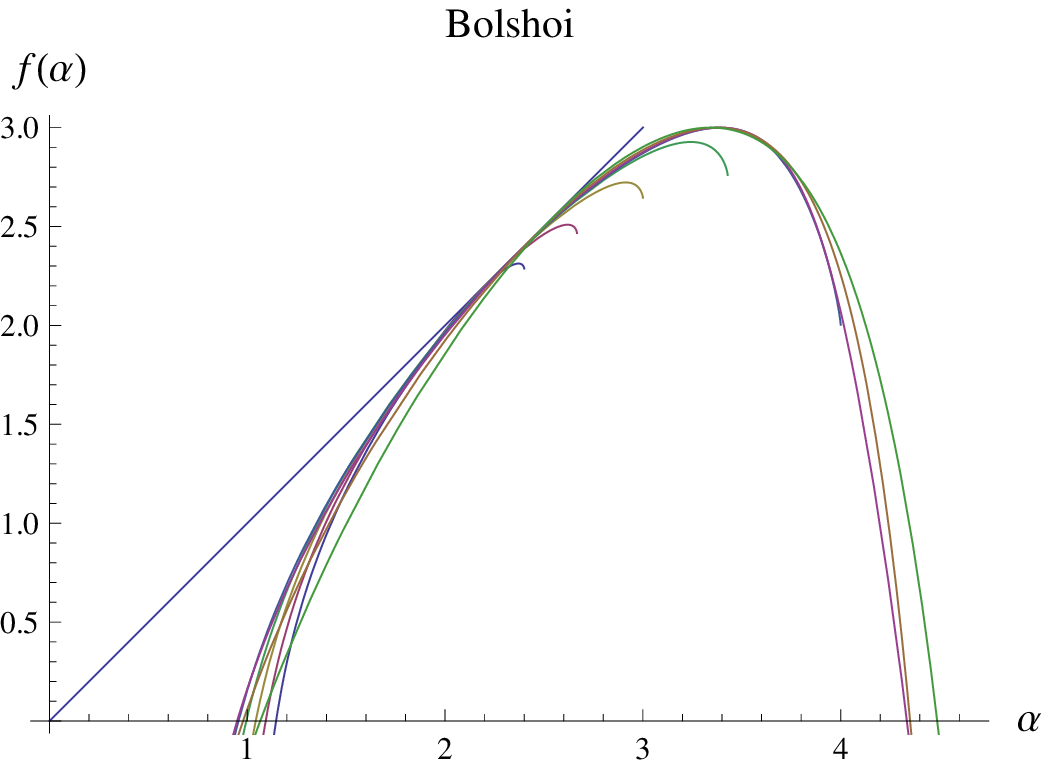}
\hspace{3mm}
\includegraphics[width=7.4cm]{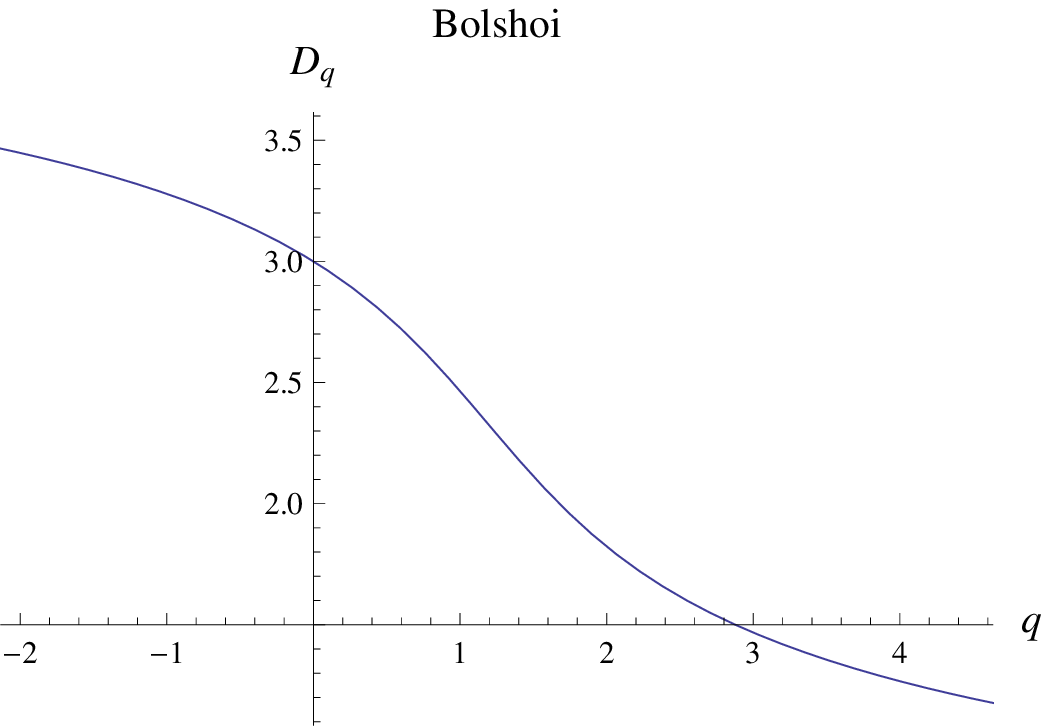}
}
\caption{(Left) Coarse multifractal spectra of the Bolshoi simulation at
  relative scales $l=2^{-13},2^{-12}, \ldots, 2^{-6}.$ (Right) R\'enyi
  dimensions at $l=2^{-7}$.}
\label{MFspec}
\end{figure}

The range between the discretization scale and the homogeneity scale in the
Bolshoi simulation is four times larger than in the Mare-Nostrum simulation,
as corresponds to the better resolution of the former.  In the Mare-Nostrum
simulation, the coarse multifractal spectra between scales 4 and 0.12 $h^{-1}$
Mpc (a factor of 32) have been shown to coincide, in the ranges where $\a$ is
defined \cite{MN}. In the Bolshoi simulation, we can proceed with the
calculation of coarse multifractal spectra to lower scales, namely, down to
0.03 $h^{-1}$ Mpc. The corresponding eight coarse multifractal spectra
(corresponding to a scale factor of 128) are plotted in Fig.\ \ref{MFspec}, on
the left-hand side. They are similar to the ones of the Mare-Nostrum
simulation and look like the typical multifractal spectrum of a self-similar
multifractal \cite{Falcon}.  The right-hand side of Fig.\ \ref{MFspec} shows
the plot of the R\'enyi dimension $D_q$, computed at the scale 2.0 $h^{-1}$
Mpc. This scale is in the middle of the interval of the three scales in Fig.\
\ref{MFspec} (left) that include the full multifractal spectrum, namely, that
include the upper-$\a$ region, corresponding to voids.

For halos, we are going to use the $q=1,2$ entropic coefficients only.  The
Bolshoi cosmic-web multifractal analysis yields R\'enyi dimensions $D_1=2.46$
and $D_2=1.82$, which are similar to those obtained from other $N$-body
simulations \cite{I4,MN}.  If we divide $D_1$ and $D_2$ by three, resulting
0.82 and 0.61, respectively, we have, approximately, the large-$r$ values of
the corresponding entropic coefficients of individual halos.  However, the
calculation of the $D_q$ of a full $N$-body simulation involves an average
and, on the other hand, the large-radius limit of the R\'enyi dimension of
spherical shells can take substantially different values for different
halos. The R\'enyi dimension $D_1$ corresponds to the point in the
multifractal spectrum such that $f(\a)=\a$, that is, to the so-called
``measure's concentrate'' (the measure is the mass) \cite{Falcon} (see also
appendix \ref{append}).  Since we can associate the region where the mass
concentrates with the set of ``typical'' halos, we deduce that $D_1$ gives a
better idea of the expected limit value of the corresponding entropic
coefficient than $D_2$ does: indeed, $q=2$ corresponds to especially
concentrated halos, so the large-radius limit of the $q=2$ R\'enyi coefficient
of spherical shells of a ``typical'' halo should have a value larger than
$D_2/3=0.61$.

Let us notice that two coarse multifractal spectra in Fig.\ \ref{MFspec}
(left) correspond to scales smaller than the discretization length, which is
$l=N^{-1/3}=2^{-11}$.  In consequence, those spectra only contain information
on the smallest values of the local dimension $\a$, that is to say, on the
densest regions.  The smallest scale, $l=2^{-13}$, is really small, namely, 31
$h^{-1}$ kpc, and the corresponding coarse multifractal spectrum hardly
reaches the point that represents the concentrate of the mass. However,
these coarse multifractal spectra, necessarily restricted to strong mass
concentrations, seem to represent these concentrations fairly well.

\subsection{Analysis of Bolshoi's halos}
\label{q-sph}

\begin{figure}
\centering{
\includegraphics[width=7.4cm]{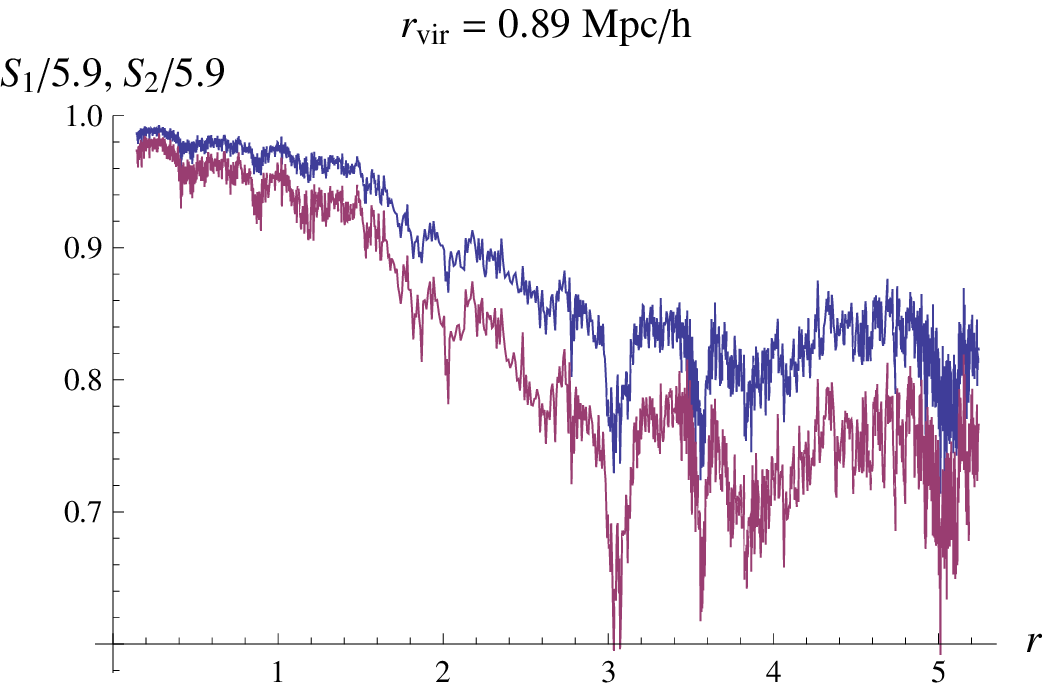}
\hspace{3mm}
\includegraphics[width=7.4cm]{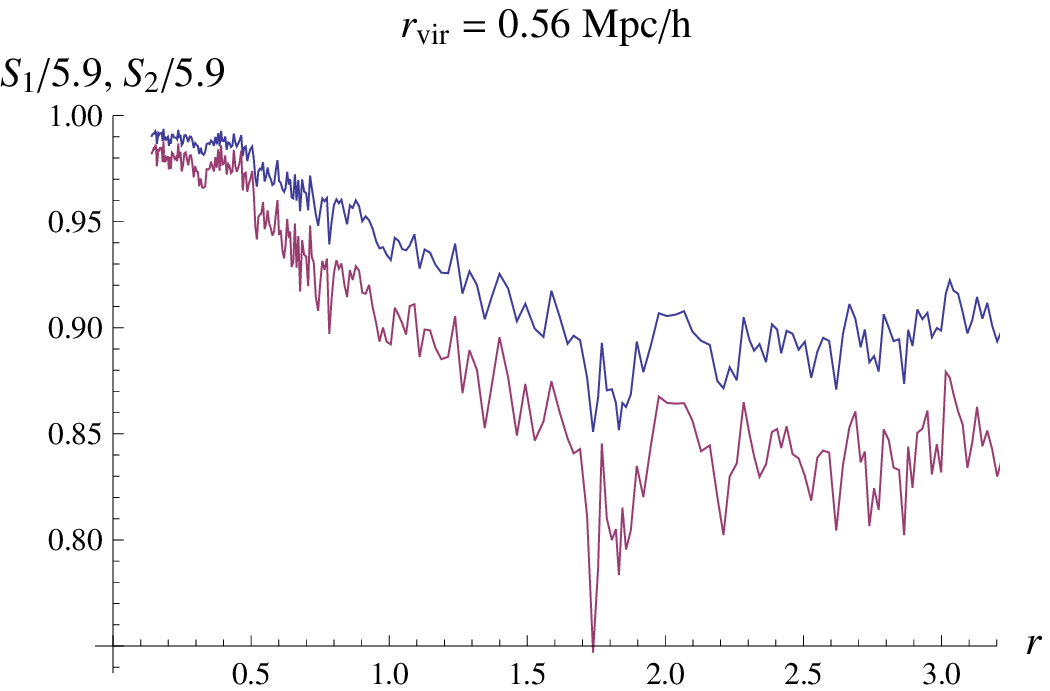}
}\\[5mm]
\centering{
\includegraphics[width=7.4cm]{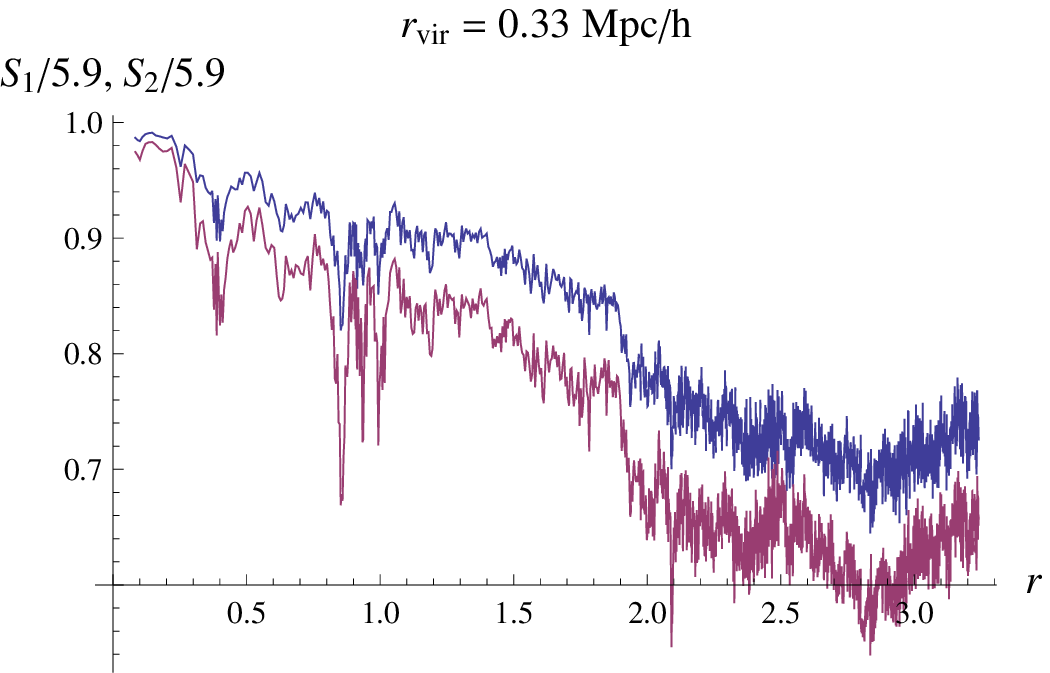}
\hspace{3mm}
\includegraphics[width=7.4cm]{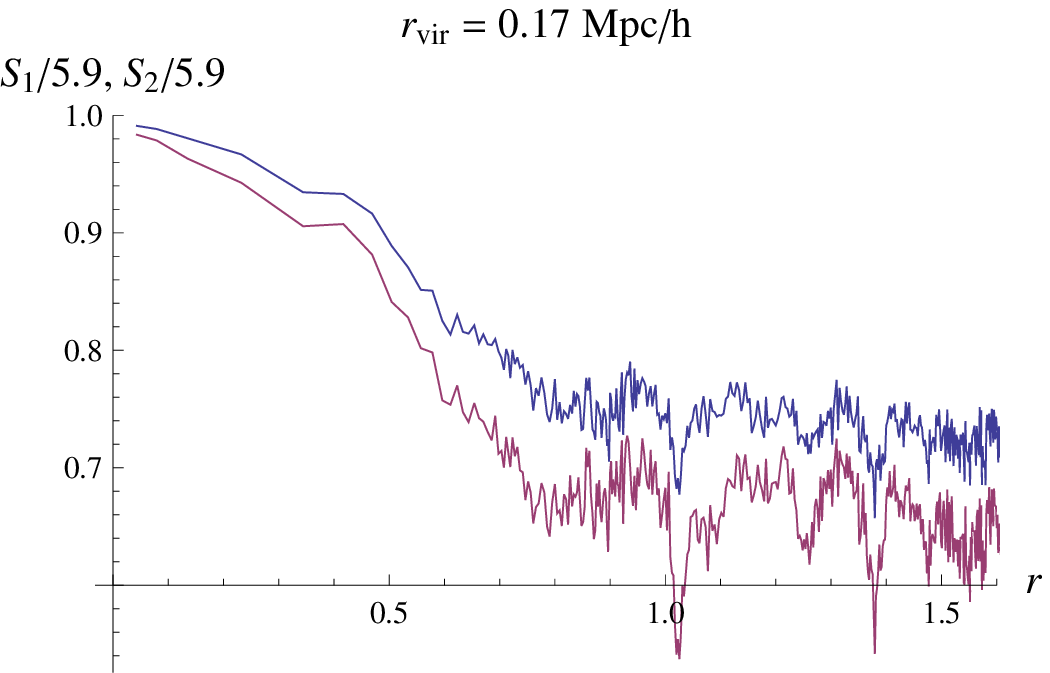}
}
\caption{Smoothness, measured by entropic coefficients,
  versus radius ($h^{-1}$ Mpc), in four quasi-spherical Bolshoi halos of
  decreasing virial radii.  The upper (blue) lines correspond to the $q=1$
  entropic coefficients and the lower (red) lines correspond to the $q=2$
  coefficients.}
\label{S12-B}
\end{figure}

The largest halo in the list of Bolshoi halos \cite{MD} has virial radius
$r_{\rm vir} = 2.14\;h^{-1}$ Mpc, and there are 269 halos with $r_{\rm vir} >
1\;h^{-1}$ Mpc.  The heaviest halos must be considered exceptional, if we take
into account that there should be at least one normal halo per homogeneity
volume. Let us quantify this concept of normality.  If we take as the
homogeneity volume a cube of $31.2$ $h^{-1}$ Mpc, which is the $1/512$
fraction of the simulation box volume, then about the 500 heaviest halos are
exceptional (unless they have approximately the same mass, which is not the
case).  Let us recall that exceptional mass concentrations give rise to
\emph{negative} fractal dimensions in the multifractal analysis of $N$-body
simulations, as explained in \cite{MN}. Therefore, we should exclude the top
500 halos, ordered by halo mass, say ($M_{\rm vir}$ = mass of bound particles
within $r_{\rm vir}$).  In fact, if we require negligible triaxiality, for
example, a ratio of minor to major axis larger than 0.85, the largest
compliant halo ranks 481th (in order of $M_{\rm vir}$).  This halo has $r_{\rm
  vir} = 0.886\;h^{-1}$ Mpc and $M_{\rm vir}= 7.47\cdot
10^{13}\;h^{-1}M_\odot$, and it is the heaviest halo that we analyze.  We have
analyzed a number of halos, calculating entropic coefficients for them, but we
select for illustration only four: the heaviest halo and other three smaller
quasi-spherical halos, with axis ratios larger than 0.9, which are distinct
halos (not subhalos), and spanning a considerable range of sizes.  The
transition from smoothness to graininess of each halo, as measured by the
$q=1$ and $q=2$ entropic coefficients, is shown in Fig.\ \ref{S12-B}.  We can
observe the progressive decrease of smoothness with radius, from values close
to one (total smoothness) to asymptotic values that correspond to the
multifractal cosmic-web structure.  As expected, there are considerable
fluctuations, due to subhalos, superposed on the decreasing trend (in spite of
the elimination of the four most populated cells of every shell).

Remarkably, nothing special happens at the virial radius in regard to
smoothness, in all the analyzed halos.  Besides, there seems to be no
proportionality or any correlation between the magnitudes of the virial
radii and the ``smoothness radii''. For example, the smoothness radius of the
second halo in Fig.\ \ref{S12-B}, with $r_{\rm vir} = 0.56\;h^{-1}$ Mpc and
$M_{\rm vir}= 1.97\cdot 10^{13}\;h^{-1}M_\odot$, is smaller than the
smoothness radius of the third one, with $r_{\rm vir} = 0.33\;h^{-1}$ Mpc and
$M_{\rm vir}= 3.01\cdot 10^{12}\;h^{-1}M_\odot$.  Furthermore, the second halo
has higher values of the asymptotic entropic coefficients than the third one.
All this suggests that the mass concentration is stronger in the third halo
than in the second halo, in spite of the fact that the third halo has smaller
virial radius and virial mass than the second one.  If we measure the strength
with the (coarse) local dimension \cite{fhalos,I4}, namely,
$$
\a = \frac{\log(M/M_0)}{\log(l/l_0)}\,,
$$
where $M$ is the mass concentrated in a volume of diameter $l$ and the values
with subscript 0 correspond to homogeneity (with $l_0=31.2\;h^{-1}$ Mpc),
and we take as $l$ twice the smoothness radius, then we obtain for the
second halo $\a=1.6$ and for the third one $\a=1.4$. The latter indicates
greater strength (let us recall that $\a \geq 0$, with 0 corresponding to
the maximum strength). 

\begin{figure}
\centering{
\includegraphics[width=7.5cm]{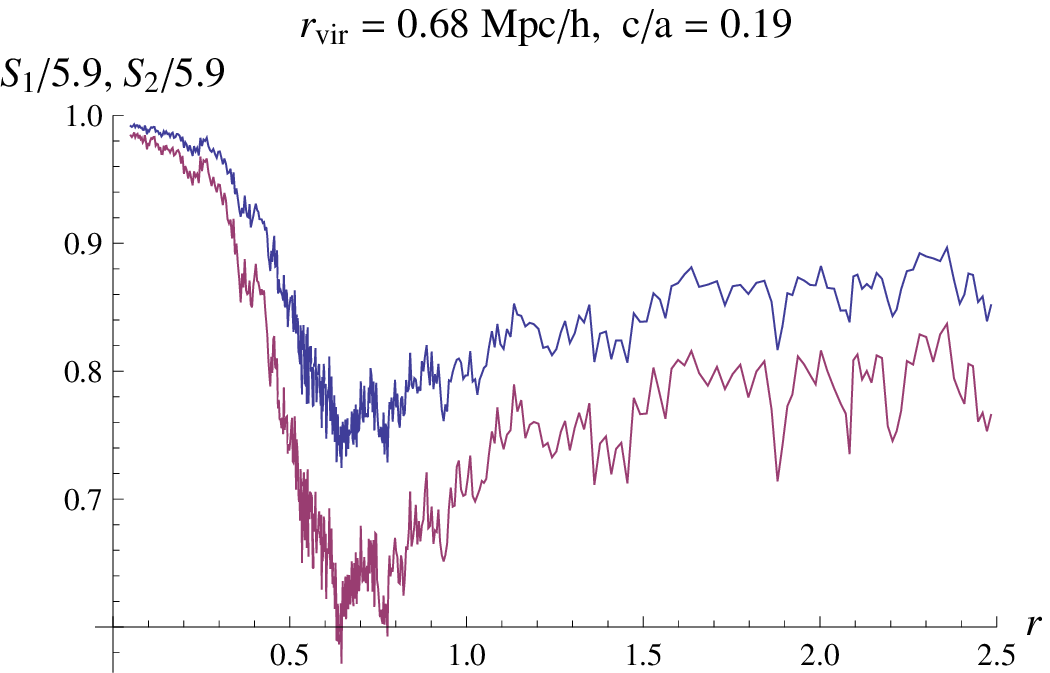}
\hspace{3mm}
\includegraphics[width=7.3cm]{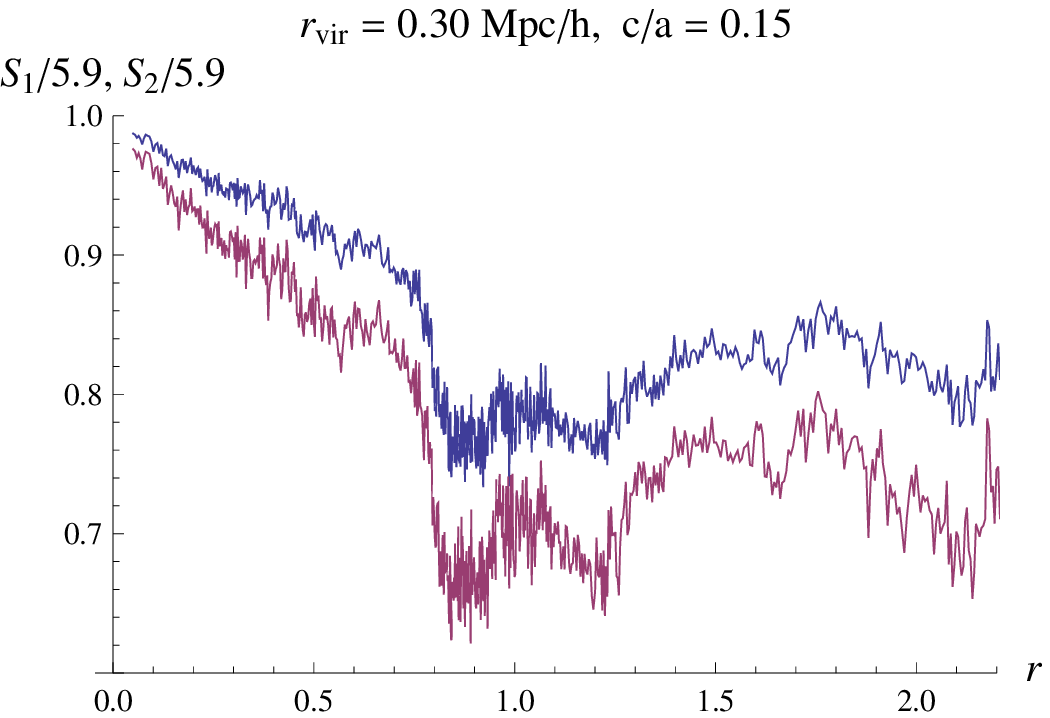}
}
\caption{Smoothness of two triaxial Bolshoi halos with small ratios of minor
  to major axes, namely, $c/a < 0.2$.}
\label{S12-c}
\end{figure}

Finally, we consider strongly triaxial halos, to determine the effect of the
associated smooth but strong variation of density inside spherical shells. We
have analyzed a number of them and we find no essential differences; namely,
the overall pattern of variation of entropic coefficients is as shown in Fig.\
\ref{S12-B}. However, the combined effect of triaxiality and insufficient mass
resolution may give rise to dips at intermediate values of $r$ in the
plots of entropic coefficients, as shown in Fig.\ \ref{S12-c}. It is natural
that the intermediate scales, where triaxiality is fully developed and is the
main anisotropic feature, are most affected. Let us notice again that any
effect of triaxiality on the entropic coefficients should vanish with
increasing mass resolution.

\subsection{Analysis of the Via Lactea II halo}

The analysis of graininess in the VL2 halo made by Zemp et al \cite{Zemp},
employing the Gini coefficient and subhalo and void frequencies, shows that
graininess steadily increases with radius.  The Gini coefficient increases
steadily from small values at small radii to $G=0.3828$ at $r=200$ kpc and to
$G=0.6193$ at $r=400$ kpc (notice that the 200 background density radius is
$r_{\mathrm{200b}}=402.1$ kpc).  The conclusion is that the outskirts of dark
matter halos have a clumpy structure \cite{Zemp}. Zemp et al do not specify
what ``the outskirts'' are, but they surely mean the regions with $r \gtrsim
400$ kpc.

\begin{figure}
\centering{
\includegraphics[width=7.4cm]{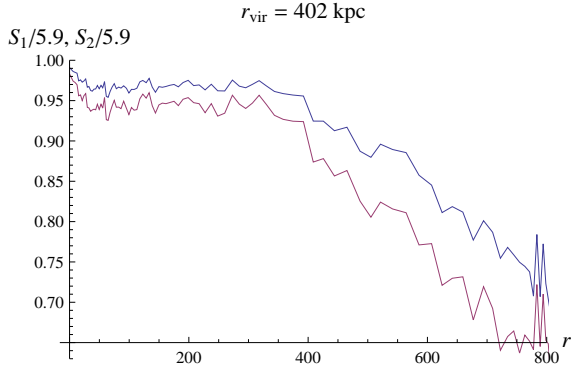}
}
\caption{Decrease of smoothness ($q=1,2$ entropic coefficients) 
with increasing radius (in kpc) 
for the VL2 halo.
}
\label{S12-VL}
\end{figure}

For our entropic analysis, we use the random subset of $100\hspace{1pt}000$
dark matter particles at redshift $z=0$ within $r=800$ kpc available at the VL
project web-page \cite{VL}.  The use of this subset might seem to reduce the
resolution of the halo, but the random selection of a subset of particles at
$z=0$ is independent of the dynamics, which corresponds to the full set of
particles.  Nevertheless, the statistical errors of smoothness measures are
larger in the reduced set of particles than in the full set. At any rate,
$100\hspace{1pt}000$ particles are sufficient for our purposes, because this
number is larger than the number of particles available for the smaller
above-analyzed Bolshoi halos; so we expect that the values of the entropic
coefficients are accurate.  Of course, the Gini coefficients obtained with
$100\hspace{1pt}000$ particles are not directly comparable with the ones
obtained with the full set of particles by Zemp et al \cite{Zemp} (our
algorithm also computes the Gini coefficients, but they do not add any
relevant information).  The $q=1$ and $q=2$ entropic coefficients shown in
Fig.\ \ref{S12-VL} confirm Zemp et al's results and give more information.

One remarkable feature of the plot of entropic coefficients (Fig.\
\ref{S12-VL}) is that these coefficients start decreasing in the first few
kiloparsecs but thence stay nearly constant up to $r=300$ kpc. This indicates
that the distribution of particles is rather smooth, on average, for small
radii and starts becoming grainy not far from $r_{\mathrm{200b}}$. Zemp et
al's quantifiers, especially, the Gini coefficient, show it as well (see
Table~1 in \cite{Zemp}). However, a detailed study of the VL2 distribution
\cite{VL2} shows that the inner density profile is ``cuspy'' (singular) and
there are many subhalos close to the center. In fact, Diemand et al \cite{VL2}
write in the abstract: ``We find hundreds of very concentrated dark matter
clumps surviving near the solar circle, \ldots The simulation reveals the
fractal nature of dark matter clustering.''

Finally, let us notice that, once that the decrease of smoothness begins in
Fig.\ \ref{S12-VL}, it continues, and the coefficients do not seem to
stabilize, although there is a slight indication that they may do so at the
end of the plot. In any event, we have to take into account that the radius
range in Fig.\ \ref{S12-VL} is considerably smaller than in the plots of Fig.\
\ref{S12-B} and it is natural that the convergence to the cosmic web take
place at larger radius.

\section{$N$-body discreteness and smoothness of halos}
\label{discret}

The smoothness of the dark matter distribution in $N$-body simulations is
presumably due to the combined effects of the diffusion by close encounters
and of the softening of strong mass concentrations.  Therefore, the smoothness
of halos depends on both the discreteness parameters, namely, $N$ (or the
discretization length), and the softening length $\e$.  The optimal choice of
softening is a debated issue \cite{KMS,SMSS,Romeo,JMB,Mer,Zhan}.  This problem
can be formulated as follows: Given a bound gravitational structure of linear
size $R$, for example, a dark matter halo represented by a set of $N$
particles, the optimal softening length $\e$ has to be determined in terms of
$N$ and $R$. There are several criteria to do it.  The deduced dependence of
the softening length $\e$ on $N$ and $R$ goes from $\e \sim RN^{-1}$, with the
classic criterion of avoiding close encounters \cite{Bin-T}, to $\e \sim
RN^{-1/3}$, that is, a softening length of the order of the discretization
length. The latter is certainly the safest choice \cite{KMS,SMSS,Romeo,JMB}.
Writing $\e \sim RN^{-\b}$, one has $1/3 \leq \b \leq 1$.  A sophisticated
statistical criterion, based on the mean integrated square error (MISE) of the
gravitational force, yields values of $\b$ that are in the range $0.2 \leq \b
\leq 0.44$ \cite{Mer,Zhan}, about the value $1/3$. The usual choice, $\e \sim
RN^{-1/2}$ \cite{DR,Power}, does not guarantee small errors in halo centers
\cite{Zhan}.

One thing to notice in all these relations is that $\e$ is proportional to
$R$. This is imposed by dimensional analysis, but only in the case of dealing
with an \emph{isolated} set of $N$ bound particles. In contrast, in
cosmological $N$-body simulations, $N$ is the total number of particles in the
simulation box, which has nothing to do with the number of particles in one
halo. In other words, we have another length parameter, namely, the size of
the simulation box $L$, so $\e$ does not have to be proportional to
$R$. Assuming that the existence of individual halos bears no relation to the
size of the simulation box and, hence, to the total number of particles $N$,
it is better to express the softening length in terms of local quantities,
namely, the discretization length, $\ell = LN^{-1/3}$, and the typical halo
size, $R$.  Now, dimensional analysis imposes no restrictions on the function
$\e(\ell,R)$.  Of course, we are also assuming that the range of halo sizes is
bounded and not too large, but this is the basic tenet of halo models
\cite{CooSh}.  Naturally, before thinking of a range of halo sizes, it is
necessary to know how to define the size of a given halo. It is normal to
choose the virial radius, derived through the spherical collapse model, but
here we consider more reasonable to resort to the smoothness properties.  In
conclusion, a reasonable value of $R$ for a given simulation should be some
average of the smoothness size of normal halos (non-exceptional halos, in the
sense discussed at the beginning of Sect.~\ref{q-sph}).

Since both the discreteness parameters $\ell$ and $\e$ are fixed once and for
all before starting the $N$-body simulation, whereas the sizes of halos belong
to the result of the simulation, it is preferable to write the relation among
the three quantities as $R(\ell,\e)$.  Ideally, this function would be weakly
dependent on the variables $\ell$ and $\e$, so these parameters would have
little influence on the size of halos, and the value of $R$ could be ascribed
to the initial conditions. In fact, there are evidences that point to a
substantial influence of $N$-body discreteness \cite{I6}.

In this regard, it would be very interesting to determine the {\em separate}
influences of $\ell$ and $\e$ on $R$. While a finite $\ell$ (a finite $N$)
introduces corrections to the mean-field Vlasov-Poisson dynamics, partially
remedied by the gravity softening, this softening also perturbs the
Vlasov-Poisson dynamics.  Joyce et al \cite{JMB} discuss the separate role of
both parameters and point out that the rigorous Vlasov-Poisson limit, $\ell
\ra 0$, is to be taken at \emph{fixed} finite $\e$, resulting in a smoothed
version of the Vlasov-Poisson equations. One may ask what $\lim_{\ell \ra
  0}R(\ell,\e)$ is. To answer this question, dimensional analysis is again
helpful, and $\lim_{\ell \ra 0}R(\ell,\e) \sim \e$. This is natural, because,
without discretization, the softening on a scale $\e$ should produce smoothing
on the same scale. In other words, if halo sizes are determined by smoothness,
these sizes must be of the order of $\e$. Therefore, the subsequent $\e \ra 0$
limit that leads to the actual Vlasov-Poisson equations makes halo sizes
vanish.  Simultaneously, the number density of halos diverges. As suggested by
Diemand et al \cite{VL2}: ``at infinite resolution one would find a long
nested series of halos within halos within halos etc.''  The simultaneous halo
size vanishing and number-density diverging imply that the relevant solutions
of the Vlasov-Poisson equations are \emph{fully} singular, that is, contain
non-isolated singularities, as corresponds to a multifractal cosmic web
structure that is present on ever decreasing scales.  Of course, the limit
$\ell \ra 0$ at \emph{fixed} finite $\e$ or, in other words, the domain of
discreteness parameters such that $\e \gg \ell$, is {\em not} studied by
cosmological $N$-body simulations.

However, information on the domain $\e \gg \ell$ is provided by the adhesion
model \cite{Gurb-Sai,Shan-Zel,GSS,Kof-Pog-Sh,WG,Kof-Pog-Sh-M,BD}. In the
adhesion model with finite viscosity $\nu$, the cosmic-web sheets, filaments
and nodes are not singular but have widths proportional to $\nu$, which plays
a regularizing role, like $\e$ does in the Vlasov-Poisson equations. Since the
adhesion model is analytically soluble in the limit $\nu \ra 0$, the exact
form of the distribution in this limit is known.  In particular, the solutions
of the adhesion model corresponding to cosmological initial conditions
contain, in the limit $\nu \ra 0$, dense sets of singular mass concentrations
of the three types: sheets, filaments and nodes \cite{V-Frisch}. By a set
being ``dense'' is meant that any volume, however small, intersects the set.
Remarkably, the formation of dense sets of singular mass concentrations is
independent of the exact type of initial power spectrum of fluctuations and is
due just to the bottom-up structure formation characteristic of CDM.  If we
associate the halo centers of the regularized Vlasov-Poisson equations with
the nodes of the adhesion model, we deduce that, when the regularization is
removed, halo sizes vanish and these zero-size ``halos'' become so prevalent
that any volume contains an infinite number of them.

\section{Summary and conclusions}
\label{discuss}

To summarize, our analysis of the smoothness or, alternately, the graininess
of halos is based on the application of robust entropic measures related to
statistical measures of inequality that are employed in economy. The entropic
coefficients that we have defined, being also related to R\'enyi dimensions,
can actually measure properties of a continuous distribution of matter,
unlike, for example, the Gini coefficient, employed by Zemp et al \cite{Zemp},
which does not have a continuum limit. The entropic coefficients of spherical
shells centered on a halo are well suited to describe the transition, as the
shell radius grows, from smoothness or mild anisotropy to the graininess or
strong anisotropy characteristic of a cosmic web structure.  The entropic
coefficients of a shell are calculated by employing counts-in-cells.  We find
it appropriate to use shells containing 1024 particles and use, for each
shell, a $8 \times 8$ mesh on the unit square of cylindrical equal-area
coordinates.

The would-be uniform distribution in an inner spherical shell of a halo is
altered by two factors: halo shape, namely, halo triaxiality, and the possible
intersection of the shell with subhalos. Both factors produce a varying
density but do not really produce non-smoothness, except if the shell
intersects singular subhalo centers. However, given that we have a limited
number of particles per shell (chosen as 1024), the statistical estimation of
entropic coefficients is subject to errors coming from both smooth and
non-smooth sources of anisotropy, which are not easily distinguishable. To
prevent anisotropy due to triaxiality, we may select quasi-spherical halos,
but triaxiality only produces trivial modifications.  At any rate, the worst
source of anisotropy is the intersection with subhalo centers. We mitigate
this effect by removing the four most populated cells of any shell (out of the
total 64 cells). However, we find that the entropic coefficients fluctuate
considerably, namely, they undergo frequent dips due to subhalos. 
Nevertheless, an average descending pattern is always clearly discernible. 

We have analyzed several halos from the Bolshoi $N$-body simulation and, also,
the Via Lactea II halo.  We find, like Zemp et al, a progressive and
essentially monotonic growth of graininess or anisotropy with growing radius
and, furthermore, we observe, in every halo, that the growth of graininess
stops at some radius and the amount of graininess stabilizes.  The radius at
which the limit graininess or anisotropy is attained marks the end of the
smooth halo and the beginning of the cosmic web structure. Indeed, the limit
values of the entropic coefficients agree with the R\'enyi dimensions of the
cosmic web, which are computed independently.  We find no proportionality or
any other definite correlation between the smoothness radii and the virial
radii of the analyzed halos, although there is a global trend of diminishing
smoothness radii with virial radii.  Besides, the smoothness radius normally
is considerably larger than the virial radius.  We propose that the smoothness
radius gives an alternative measure of halo size that may be more convenient
in some regards. Of course, one must not necessarily conclude that smoothness
is independent of dynamical relaxation to stable states (``virialization'')
but just that the virial radius may not be an adequate measure of a stable
state and also that relaxation may be influenced by $N$-body discreteness
effects.

In fact, the smoothness of halos in $N$-body simulations can be mainly due to
discreteness effects, as indicated by our qualitative analysis.  A
quantitative analysis of the effects of $N$-body discreteness demands a deeper
understanding of the influence of the discreteness parameters or, in other
words, of the nature of the function $R(\ell,\e)$ that describes the size of
halos in terms of the fundamental discreteness parameters, namely, the
discretization length $\ell$ and the softening length $\e$.  Nevertheless, we
argue that the removal of these parameters in a physically meaningful way may
lead to the vanishing of halo sizes, while the halo number density diverges.

\appendix
\section{Appendix: multifractal analysis}
\label{append}

Coarse multifractal analysis is appropriate, in general, for physical examples
and the results of simulations \cite{Falcon}. For a distribution of particles,
the mass (the ``measure'') is discretized and the method comes down to an
elaboration of the counts-in-cells statistics that is common in the analysis
of $N$-body simulations.

Let us assume that a mesh of cells is placed in the sample region, that is,
for our purposes, either the full simulation box or the unit square that
corresponds to a spherical shell of a halo (Sect.\ \ref{square}).  Fractional
statistical moments are defined by counts in cells as
\begin{equation}
M_q = \sum_i \left(\frac{n_i}{N}\right)^{q} = 
\sum_{n>0} N(n)\left(\frac{n}{N}\right)^{q},
\label{Mq}
\end{equation}
where the index $i$ refers to non-empty cells, $n_i$ is the number of
particles in the cell $i$, $N= \sum_i n_i$ is the total number of particles,
and $N(n)$ is the number of cells with $n$ particles.  The second expression
involves a sum over cell populations that is more useful than the sum over
individual cells when most cells are scarcely populated.  $M_0$ is the number
of non-empty cells and $M_1$ is normalized to 1.

In regular distributions, the mass (number of particles) contained in any cell
must be proportional to the cell's volume, $v$, for sufficiently small
$v$. Therefore, $M_q \sim v^{q-1}$.  This does not apply to singular
distributions, but they can be such that their $q$-moments are non-trivial
power laws of $v$ in the $v \ra 0$ limit. So one can define, for a singular
distribution, the non-trivial exponents
\begin{equation}
\tau(q) = 3\lim_{v\ra 0}\frac{\log M_q}{\log v}\,,\;q \in \mathbb{R}\,,
\label{tauq}
\end{equation}
provided that the limit exists.
Such a distribution is called multifractal.
Of course, the numerical evaluation of the limit in Eq.~(\ref{tauq}) is not
feasible and one must be satisfied with finding a constant value of the
quotient for sufficiently small $v$, that is, in a range of negative values of
$\log v$ (a range of scales).  In fact, the exponent is normally defined as
the slope of the plot of $\log M_q$ versus $\log v$, and its value is found by
numerically fitting that slope.

A multifractal is also characterized by its {\em local} dimensions. The local
dimension $\a$ at the point $\bm{x}$ expresses the asymptotic power-law form
of the mass growth from that point outwards, $m(\bm{x},r) \sim
r^{\a(\bm{x})}$, and defines the ``strength'' of the corresponding
singularity. Actually, singularities correspond to $\a < 3$, whereas points
with $\a \geq 3$ are regular.  Every set of points with a given local
dimension $\a$ constitutes a fractal set with a dimension that depends on
$\a$, namely, $f(\a)$.  In terms of $\tau(q)$, the spectrum of local
dimensions is given by
\begin{equation}
\a(q)= \tau'(q)\,,\quad q \in \mathbb{R}\,,
\label{aq}
\end{equation}
and the spectrum of fractal dimensions $f(\a)$ is given by the Legendre
transform
\begin{equation}
f(\a) = q\,\a - \tau(q)\,.
\label{fa}
\end{equation}
Self-similar multifractals have a typical spectrum of fractal dimensions that
spans an interval $[\a_{\mathrm{min}},\a_{\mathrm{max}}]$, is convex from
above, and fulfills $f(\a) \leq \a$.  Furthermore, the equality $f(\a)=\a$ is
reached at one point, such that $q=1$ in Eq.~(\ref{fa}): note that
Eq.~(\ref{tauq}) gives $\tau(1)=0$.  The corresponding set of singularities
contains the bulk of the mass and is called the ``mass concentrate.''

As said above, the convergence to the limit in Eq.~(\ref{tauq}) must take
place in a range of small values of $v$. Naturally, $v$ must be small in
comparison with the homogeneity volume $v_0$, which is the smallest volume
such that the mass fluctuations in it are small and approximately
Gaussian. Therefore, we define, for a given cell size $v$, the {\em coarse}
exponent
\begin{equation}
\tau(q) = 3\frac{\log (M_q/v_0^{q-1})}{\log (v/v_0)}\,.
\label{ctauq}
\end{equation}
For cell sizes larger than $v_0$, $M_q \sim v^{q-1}$ and $\tau(q) = 3(q-1).$
The coarse exponent $\tau$ depends on both $v$ and $v_0$, but it must depend
mildly on the latter.  Nevertheless, this dependence on $v_0$ is generally
non-negligible: if one just sets $v_0$ to 1, that is to say, to the total
volume, as often done, the coarse exponents may be so inaccurate that the
multifractal scaling is spoiled. In other words, if $v_0$ is not included, the
available range of $v$ may not be long enough for us to obtain reliable values
of the functions $\tau(q)$ and $f(\a)$.  One can estimate $v_0$ as, for
example, the coarse-graining scale such that the mass fluctuations are smaller
than a given fraction, say, 10\%.

As a complement to the multifractal spectrum $f(\a)$, it is useful to define
the spectrum of R\'enyi dimensions 
\begin{equation}
D_q= \frac{\tau(q)}{q-1}\,,
\label{Dq}
\end{equation}
because they have an information-theoretic meaning. Indeed, they are related
to R\'enyi entropies \cite{Renyi}; namely, they express the power-law behavior
of the R\'enyi entropies of the coarse distribution in the limit $v \ra 0$:
$$
D_q = \lim_{v \to 0} \frac{3\,S_q(\{p_i\})}{-\log_2 v}.
$$ 
R\'enyi entropies, in general, measure the lack of information or the
uncertainty of a probability distribution. In the case of a discrete
distribution of particles, they measure the uncertainty in the choice of $q$
particles (when $q$ is a positive integer).  The dimension of the mass
concentrate $\a_1 = f(\a_1) = D_1$ is also called the entropy dimension.
$D_0$ coincides with the maximum value of $f(\a)$ andq with the box-counting
dimension of the distribution's support, while $D_2 = \tau(2)$ is the
correlation dimension.  In the homogeneous regime, $\tau(q) = 3(q-1)$ and
$D_q= 3$ for any $q$.  In a uniform fractal (a {\em unifractal} or {\em
  monofractal}) $D_q$ is also constant but smaller than three. In general,
$D_q$ is a non-increasing function of $q$.


\begin{thebibliography}{99}

\bibitem{Zel} 
Ya.B. Zeldovich,
\emph{Gravitational instability: An approximate theory for large density
perturbations, Astron.\ \& Astrophys.} {\bf 5} (1970) 84--89.

\bibitem{Gurb-Sai} 
S.N. Gurbatov and A.I. Saichev, 
\emph{Probability Distribution and Spectra of Potential Turbulence,
Radiophys.\ Quant.\ Electr.}
\textbf{27} (1984) 303--313.

\bibitem{Shan-Zel} 
S.F. Shandarin and Ya.B. Zel'dovich,
\emph{The large-scale structure of the universe: Turbulence, intermittency,
  structures in a self-gravitating medium,
Rev. Mod. Phys.}
{\bf 61} (1989) 185--220.

\bibitem{GSS}
S.N. Gurbatov, A.I. Saichev and S.F. Shandarin, 
\emph{Large-scale structure of the Universe. The Zeldovich approximation 
and the adhesion model, 
Phys.\ Usp.} {\bf 55} (2012) 223--249.

\bibitem{EJS}
J. Einasto, M. J\~oeveer and E. Saar, 
\emph{Structure of superclusters and supercluster formation, 
MNRAS} {\bf 193} (1980) 353--375.

\bibitem{ZES}
Ya.B. Zeldovich, J. Einasto and S.F. Shandarin,
\emph{Giant Voids in the Universe, Nature} {\bf 300} (1982) 407--413.

\bibitem{Ge-Hu}
M.J. Geller and J.P. Huchra,
\emph{Mapping the Universe, Science}
{\bf 246} (1989) 897--903.

\bibitem{Kof-Pog-Sh}
L. Kofman, D. Pogosyan and S.F. Shandarin,
{\em Structure of the universe in the two-dimensional model of adhesion,
MNRAS}  
{\bf 242} (1990) 200--208.

\bibitem{WG} 
D. H. Weinberg and J. E. Gunn,
\emph{Large-scale Structure and the Adhesion Approximation,
MNRAS}  
{\bf 247} (1990) 260--286.

\bibitem{Kof-Pog-Sh-M}
L. Kofman, D. Pogosyan, S.F. Shandarin and A.L. Melott,
{\em Coherent structures in the universe and the adhesion model, 
The Astrophysical Journal} 
{\bf 393} (1992) 437--449.

\bibitem{CooSh}
A. Cooray and R. Sheth,
\emph{Halo models of large scale structure,
Phys. Rep.} {\bf 372} (2002) 1--129.

\bibitem{MoW}
H.J. Mo and S.D.M. White, 
\emph{An analytic model for the spatial clustering of dark matter haloes, 
MNRAS} {\bf 282} (1996) 347--361.

\bibitem{CMP}
P. Catelan, S. Matarrese and C. Porciani, 
\emph{On the Spatial Distribution of Dark Matter Halos,
Astrophys.\ J.}  {\bf 502} (1998) L1--L4.

\bibitem{Sheth-Tor} R.K. Sheth, and G. Tormen, \emph{Large-scale bias and the
    peak background split, MNRAS} {\bf 308} (1999) 119--126.

\bibitem{Power}
C. Power et al,
\emph{The inner structure of ΛCDM haloes - I. A numerical convergence study,
MNRAS} {\bf 338} (2003) 14--34.

\bibitem{Hayashi}
E. Hayashi et al,
\emph{The inner structure of ΛCDM haloes - II. Halo mass profiles and low
  surface brightness galaxy rotation curves,
MNRAS} {\bf 355} (2004) 794--812.

\bibitem{Navarro}
J.F. Navarro et al,
\emph{The inner structure of ΛCDM haloes - III. Universality and asymptotic
  slopes,
MNRAS} {\bf 349} (2004) 1039--1051.

\bibitem{Mandel} B.B. Mandelbrot, \emph{The fractal geometry of nature} (rev.\
  ed.\ of: \emph{Fractals}, 1977), W.H. Freeman and Company (1983).

\bibitem{V-Frisch} M. Vergassola, B. Dubrulle, U. Frisch and A. Noullez,
\emph{Burgers' equation, Devil's staircases and the mass distribution for
  large-scale structures, 
Astron.\ \& Astrophys.} {\bf 289} (1994) 325--356.

\bibitem{Kturb} J. Gaite, \emph{A non-perturbative Kolmogorov turbulence
    approach to the cosmic web structure, Europhys. Lett.} {\bf 98} (2012)
  49002.

\bibitem{Rigo} G. Rigopoulos, \emph{The adhesion model as a field theory for
    cosmological clustering, JCAP} {\bf 1} (2015) 014.

\bibitem{Bou-M-Parisi} J.P. Bouchaud, M. M\'ezard and G. Parisi,
\emph{Scaling and intermittency in Burgers turbulence,
Physical Review E} {\bf 52} (1995) 3656--3674.

\bibitem{Pietronero}
L. Pietronero, \emph{The fractal structure of the universe: Correlations of
  galaxies and clusters and the average mass density, 
Physica A} {\bf 144} (1987) 257--284.

\bibitem{Jones} B.J. Jones, V. Mart\'{\i}nez, E. Saar and J. Einasto, \emph{
    Multifractal description of the large-scale structure of the universe,
    Astrophys.\ J.}  {\bf 332} (1988) L1--L5.

\bibitem{Bal-Schaf} R. Balian and R. Schaeffer, \emph{Galaxies: Fractal
    dimensions, counts in cells, and correlations, Astrophys.\ J.}  {\bf 335}
  (1988) L43--L46.

\bibitem{Jones-RMP} B.J. Jones, V. Mart\'{\i}nez, E. Saar and V. Trimble,
  \emph{Scaling laws in the distribution of galaxies, Rev. Mod. Phys.}  {\bf
    76} (2004) 1211--1266.

\bibitem{Valda} R. Valdarnini, S. Borgani and A. Provenzale,
  \emph{Multifractal properties of cosmological N-body simulations,
    Astrophys.\ J.}  {\bf 394} (1992) 422--441.

\bibitem{Colom} S. Colombi, F.R. Bouchet and R. Schaeffer,	
\emph{Multifractal analysis of a cold dark matter universe,
Astron.\ \& Astrophys.} {\bf 263} (1992) 1.

\bibitem{Yepes} G. Yepes, R. Dom{\'\i}nguez-Tenreiro and H.P.M. Couchman, 
\emph{The scaling analysis as a tool to compare N-body simulations with
  observations --- Application to a low-bias cold dark matter model,
  Astrophys.\ J.} 
  {\bf 401} (1992) 40--48.

\bibitem{fhalos}
J. Gaite, 
\emph{The fractal distribution of haloes,
Europhys. Lett.} {\bf 71} (2005) 332--338.

\bibitem{I4}  J. Gaite,	
\emph{Halos and voids in a multifractal model of cosmic structure, 
Astrophys.\ J.}
{\bf 658} (2007) {11--24}.

\bibitem{voids} J. Gaite, \emph{Statistics and geometry of cosmic voids, JCAP}
  {\bf 11} (2009) 004.

\bibitem{MN}
J. Gaite,
\emph{Fractal analysis of the dark matter and gas distributions in the
  Mare-Nostrum universe, JCAP} {\bf 3} (2010) 006.

\bibitem{ChaCa}
C.A. Chac\'on-Cardona and R.A. Casas-Miranda,
\emph{Millennium simulation dark matter haloes: multifractal and lacunarity
  analysis and the transition to homogeneity, MNRAS} {\bf 427} (2012)
2613--2624. 

\bibitem{I6}
J. Gaite, \emph{Halo Models of Large Scale Structure and Reliability of
Cosmological N-Body Simulations, Galaxies} {\bf 1} (2013) 31--43. 
 
\bibitem{Bolshoi} A.A. Klypin, S. Trujillo-Gomez and J. Primack,
\emph{Dark Matter Halos in the Standard Cosmological Model: Results from the
  Bolshoi Simulation, 
Astrophys. J.} {\bf 740} (2011) 102.

\bibitem{MD}
K. Riebe et al, 
\emph{The MultiDark Database: Release of the Bolshoi and MultiDark
  Cosmological Simulations}
\href{http://arxiv.org/abs/1109.0003}
{[{\tt arXiv:1109.0003}]}.

\bibitem{halo-find}
A. Knebe et al, 
\emph{Structure finding in cosmological simulations: the state of affairs,
MNRAS} {\bf 435} (2013) 1618--1658.

\bibitem{CDM_PNAS}
D.H. Weinberg et al,
\emph{Cold dark matter: controversies on small scales}
\href{http://arxiv.org/abs/1306.0913}
{[{\tt arXiv:1306.0913}]}.

\bibitem{Zemp}
M. Zemp et al, 
\emph{The graininess of dark matter haloes, 
MNRAS} {\bf 394} (2009) 641--659.

\bibitem{VL2}
J. Diemand et al,
\emph{Clumps and streams in the local dark matter distribution, 
Nature} {\bf 454} (2008) 735--738.

\bibitem{DR} W. Dehnen and J.I. Read,
\emph{$N$-body simulations of gravitational dynamics, 
Eur. Phys. J. Plus} {\bf 126} (2011) 55:1--55:28.

\bibitem{KMS} 
B. Kuhlman, A.L. Melott and S.F. Shandarin,
\emph{A Test of the Particle Paradigm in N-Body Simulations,
Astrophys.\ J.} {\bf 470} (1996)  L41.

\bibitem{SMSS} R.J. Splinter, A.L. Melott, S.F. Shandarin and Y. Suto,
\emph{Fundamental Discreteness Limitations of Cosmological $N$-Body Clustering
Simulations,
Astrophys.\ J.} {\bf 497} (1998) 38--61. 

\bibitem{Romeo} 
A.B. Romeo, O. Agertz, B. Moore and J. Stadel,
\emph{Discreteness Effects in ΛCDM Simulations: A Wavelet-Statistical View,
Astrophys.\ J.} {\bf 686} (2008) 1--12. 

\bibitem{JMB}
M. Joyce, B. Marcos and T. Baertschiger,
\emph{Towards quantitative control on discreteness error in the non-linear
  regime of cosmological N-body simulations, MNRAS} 
{\bf 394} (2009) 751--773.

\bibitem{Shanda-JCAP} S.F. Shandarin,
\emph{The multi-stream flows and the dynamics of the cosmic web, JCAP} {\bf
  15} (2011) 005. 

\bibitem{Abel} T. Abel, O. Hahn and Kaehler,
\emph{Tracing the dark matter sheet in phase space, MNRAS}  
{\bf 427} (2012) 61--76.

\bibitem{Ney} M. Neyrinck, 
\emph{Origami constraints on the initial-conditions arrangement of dark-matter
  caustics and streams, MNRAS}  
{\bf 427} (2012) 494--501.

\bibitem{S3} V. Sahni, B.S. Sathyaprakash and S.F. Shandarin,
\emph{Shapefinders: A New Shape Diagnostic for Large-Scale Structure,
The Astrophysical Journal} 
{\bf 495} (1998) L5.

\bibitem{S4} J.V. Sheth, V. Sahni, S.F. Shandarin and B.S. Sathyaprakash,
\emph{Measuring the geometry and topology of large-scale structure using
  SURFGEN: 
methodology and preliminary results, MNRAS}
{\bf 343} (2003) 22--46.

\bibitem{vW-Sch} R. van de Weygaert and W. Schaap, \textit{The Cosmic Web:
    Geometric Analysis}, in 
\emph{Data Analysis in Cosmology (Valencia)},
  eds. V. Mart\'{\i}nez, E. Saar, E. Mart\'{\i}nez-Gonzalez,
  M.J. Pons-Border\'{\i}a, 
Lecture Notes in Physics, vol. 665, Springer-Verlag (2009) 291--413.

\bibitem{mathworld}
E.W. Weisstein, 
{\em Cylindrical Equal-Area Projection,}
from MathWorld --- A Wolfram Web Resource
\href{http://mathworld.wolfram.com/CylindricalEqual-AreaProjection.html}
{[{\tt http://mathworld.wolfram.com/CylindricalEqual-AreaProjection.html}]}.
\bibitem{Ul}
A. Ullah and D.E.A. Giles (Eds.), 
\emph{Handbook of Applied Economic
Statistics,} CRC Press (1998). 

\bibitem{Renyi} A. R\'enyi, 
\emph{Calcul des probabilit\'es}, Dunod, Paris (1966).

\bibitem{Falcon} K. Falconer, 
\emph{Fractal Geometry (Second Edition)}, John Wiley and Sons, Chichester UK,
(2003), {Chapter 17}.

\bibitem{VL}
J. Diemand et al,
{\em The via lactea project}
\href{http://www.ics.uzh.ch/~diemand/vl/}
{[{\tt http://www.ics.uzh.ch/$\sim$diemand/vl}]}.

\bibitem{Mer}
D. Merritt,
\emph{Optimal Smoothing for $N$-Body Codes, 
Astronom.\ J.}  
{\bf 111} (1996) 2462--2464. 

\bibitem{Zhan}
Hu Zhan,
\emph{Optimal Softening for $N$-Body Halo Simulations, Astrophys.\ J.} 
{\bf 639} (2006) 617--620. 

\bibitem{Bin-T}
J.~Binney and S.~Tremaine, \emph{Galactic Dynamics (Second Edition)},
Princeton University Press (2008), pg.~33.

\bibitem{BD} T. Buchert and A. Dom{\'\i}nguez, \emph{Adhesive gravitational
    clustering, Astron.\ \& Astrophys.}  {\bf 438} (2005) 443--460.

\end{thebibliography}
\end{document}